\documentclass[aps,twocolumn,nofootinbib,prd]{revtex4}
\usepackage{graphicx}
\usepackage{amsfonts}
\usepackage{amssymb}
\usepackage{amsbsy}
\usepackage{amsmath}
\usepackage{latexsym}
\usepackage{natbib}
\usepackage{bm}
\usepackage{accents}
\usepackage{color}
\usepackage{hyperref}
\usepackage{float}
\newcommand{\udt}[3]{#1^{#2}_{\phantom{#2}#3}}
\newcommand{\udut}[4]{#1^{#2\phantom{#3}#4}_{\phantom{#2}#3\phantom{#4}}}

\newcommand{\dut}[3]{#1_{#2}^{\phantom{#2}#3}}
\newcommand{\dudt}[4]{#1_{#2\phantom{#3}#4}^{\phantom{#2}#3}}
\newcommand{\lc}[1]{\accentset{\circ}{#1}}%Levi-Civita connection
\begin{document}
\title{Geodesic Congruences and a Collapsing Stellar Distribution in $f(T)$ Theories}

\author{Soumya Chakrabarti\footnote{soumya.chakrabarti@saha.ac.in}}
\affiliation{Theory Division,\\
Saha Institute of Nuclear Physics, \\
Kolkata 700064, West Bengal, India
}

\author{Jackson Levi Said\footnote{jackson.said@um.edu.mt}}
\affiliation{Institute of Space Sciences and Astronomy, University of Malta, Malta}
\affiliation{Department of Physics, University of Malta, Malta}

\date{\today}

\begin{abstract}
Teleparallel Gravity (TG) describes gravitation as a torsional- rather than curvature-based effect. As in curvature-based constructions of gravity, several different formulations can be proposed, one of which is the Teleparallel equivalent of General Relativity (TEGR) which is dynamically equivalent to GR. In this work, we explore the evolution of a spatially homogeneous collapsing stellar body in the context of two important modifications to TEGR, namely $f(T)$ gravity which is the TG analogue of $f(R)$ gravity, and a nonminimal coupling with a scalar field which has become popular in TG for its effects in cosmology. We explore the role of geodesic deviation to study the congruence of nearby particles in lieu of the Raychaudhuri equation. We find $f(T)$ models that satisfy the null energy condition and describe interesting collapse profiles. In the case of a nonminimally coupled scalar field, we also find potential collapse models with intriguing scalar field evolution profiles.
\end{abstract}

\maketitle

\section{Introduction}
The $\Lambda$CDM cosmological model is demonstrated by overwhelming observational evidence in describing the evolution of the Universe at all scales \cite{misner1973gravitation,Clifton:2011jh} which is achieved by the inclusion of matter beyond the standard model of particle physics. This appears as dark matter which stabilizes galactic structures \cite{Baudis:2016qwx,Bertone:2004pz} in the form of cold dark matter particles, and dark energy which is represented by the cosmological constant \cite{Peebles:2002gy,Copeland:2006wr} and produces late-time accelerated cosmic expansion \cite{Riess:1998cb,Perlmutter:1998np}. On the other hand, despite great efforts, there continue to remain outstanding internal problems in the cosmological constant \cite{RevModPhys.61.1}, as well as no direct observations of dark matter particles \cite{Gaitskell:2004gd}.

In addition to these issues, the effectiveness of the $\Lambda$CDM model has also been called into question in recent years. Primarily, the core critique is rooted in the so-called $H_0$ tension problem which quantifies the inconsistency between the measured \cite{Riess:2019cxk,Wong:2019kwg} and predicted \cite{Aghanim:2018eyx,Ade:2015xua} values of $H_0$ between early- and late-time observations. Measurements made on the tip of the red giant branch (TRGB, Carnegie-Chicago Hubble Program) have reported a lower tension \cite{Freedman:2019jwv}, but ultimately the problem may be clarified by future observations from more exotic sources such as gravitational wave astronomy with observatories such as the LISA mission \cite{Baker:2019nia,2017arXiv170200786A} which have already shown an ability to tackle these measurements \cite{Graef:2018fzu,Abbott:2017xzu}.

At its core, the $\Lambda$CDM model is made up of modifications to the matter section. However, modifications to the gravitational section may also provide a suitable explanation to some of the outstanding problems in modern cosmology. This has come in several forms with modifications to general relativity (GR) (see Ref.\cite{Clifton:2011jh,Capozziello:2011et} and references therein) being the main flavor in which exotic gravity enters cosmology such as in extended theories of gravity \cite{Sotiriou:2008rp,Faraoni:2008mf,Capozziello:2011et}. Collectively, these models of gravity are bourne out of GR through the common mechanism by which gravitation is expressed, i.e. the curvature associated with the Levi-Civita connection \cite{misner1973gravitation}. While the metric quantifies the amount of geometric deformation that gravity produces, its the connection which selects curvature as the property over which this is expressed \cite{BeltranJimenez:2019tjy,nakahara2003geometry}. This is not the only choice in this regard, while retaining the metricity condition, torsion has become an increasingly popular choice for constructing cosmologically motivated theories of gravity \cite{Aldrovandi:2013wha,Cai:2015emx,Krssak:2018ywd}.

Teleparallel Gravity (TG) embodies the collection of theories of gravity in which gravity is expressed as geometric torsion through the Weitzenb\"{o}ck connection \cite{Weitzenbock1923}. This connection is torsion-ful and curvatureless, whereas the Levi-Civita connection is curvature-ful and torsion-less. All curvature quantities calculated using the Weitzenb\"{o}ck connection (instead of the Levi-Civita connection) naturally vanishes irrespective of the metric components. Immediately, we can confront the Einstein-Hilbert action whose Lagrangian is simply the Ricci scalar, $\mathring{R}$ (over-circles represent quantities calculated with the Levi-Civita connection), which produces the GR field equations. The identical dynamical equations can be arrived at in TG by replacing this Lagrangian with its so-called torsion scalar, $T$, counterpart. This is the so-called \textit{Teleparallel equivalent of General Relativity} (TEGR), and differs from GR only at the level of Lagrangian by a total divergence quantity, $B$ (boundary term). 

In TEGR, the boundary term encapsulates the fourth-order corrections which appear in the action to result in a covariant theory (due to the second-order derivatives in the Einstein-Hilbert action). The impact of this feature is that extensions to TEGR will have a meaningful difference to their Levi-Civita connection counterparts. Principally this will mean that TG will have a broader range of modified theories in which the dynamical equations are second-order rather than the limits imposed by the Lovelock theorem in theories of gravity based on the Levi-Civita connection \cite{Lovelock:1971yv,Gonzalez:2015sha,Bahamonde:2019shr}. As an aside, TG also has several interesting properties such as its likeness to Yang-mills theory \cite{Aldrovandi:2013wha} giving it an added particle physics dimension, its possible definition of a gravitational energy-momentum tensor \cite{Blixt:2018znp,Blixt:2019mkt}, and that it does not require the introduction of a Gibbons--Hawking--York boundary term in order to produce a well-defined Hamiltonian formulation \cite{Cai:2015emx} making it more regular than GR. Moreover, TG can be constructed without the necessity of the weak equivalence principle \cite{Aldrovandi:2004fy} unlike GR.

Keeping to the same reasoning as $f(R)$ gravity \cite{Sotiriou:2008rp,Faraoni:2008mf,Capozziello:2011et}, TEGR can be arbitrarily generalized to produce $f(T)$ gravity \cite{Ferraro:2006jd,Ferraro:2008ey,Bengochea:2008gz,Linder:2010py,Chen:2010va,Bahamonde:2019zea}. Due to the weakened Lovelock theorem in TG, this will be a generally second-order theory of gravity which is a notable difference to its $f(\mathring{R})$ gravity analogue. $f(T)$ gravity has shown a number of promising results in recent years, in terms of cosmology \cite{Cai:2015emx,Nesseris:2013jea,Farrugia:2016qqe}, galactic physics \cite{Finch:2018gkh} as well as in solar system scale phenomenology \cite{Farrugia:2016xcw,Iorio:2012cm,Ruggiero:2015oka,Deng:2018ncg}. However, to fully embrace the possibility of limiting to $f(\mathring{R})$ cosmological models, we must consider the fuller $f(T,B)$ theory of gravity \cite{Bahamonde:2015zma,Capozziello:2018qcp,Bahamonde:2016grb,Paliathanasis:2017flf,Farrugia:2018gyz,Bahamonde:2016cul,Bahamonde:2016cul,Wright:2016ayu}, in which $f(\mathring{R})=f(-T+B)$. $f(T,B)$ gravity is an interesting theory due to the decoupling between the second-order torsion scalar and fourth-order boundary term contributions. On the other hand, extensions of TG in which matter is nonminimally coupled have also gained in popularity in recent years \cite{Bahamonde:2017ifa,Harko:2014aja,Pace:2017aon,Farrugia:2016pjh,Saez-Gomez:2016wxb,Momeni:2014jja,Nassur:2015zba,Pace:2017dpu}. These have produced interesting results in cosmology and for compact objects.

In this work, we consider the geodesic deviation of test particles in these modifications of TG. This kind of study can be very informative for investigating particular models in these extensions of TEGR. We also study the Raychaudhuri equation with a focus on the energy conditions that result from these models. Finally, we apply the results of this work to the homogeneous collapse of stellar matter. The paper is broken into the following section, first the Raychaudhuri equation is reviewed in \S.~\ref{ray_eq} while TG is briefly discussed in \S.~\ref{TG_background}. In \S.~\ref{collapse}, we explore a collapsing stellar mass distribution and the associated energy conditions for the models being investigated, while we do this again for a particularly interesting nonmimally coupled scalar-tensor model in \S.~\ref{non_min_TG}. Finally in \S.~\ref{conc} we discuss the main results and conclusions from this work. In all that follows, Latin indices are used to refer to tangent space coordinates, while Greek indices refer to general manifold coordinates.

\section{\label{ray_eq} The Raychaudhuri Equations}
The Raychaudhuri equation \cite{1955PhRv...98.1123R} offers an efficient avenue by which the tendency for nearby geodesics to converge in a gravitational system can be concisely described. One of the most notable impacts of this scheme was in the focusing theorem. Geodesic focusing \cite{hawking_ellis_1973} is a natural consequence of the Raychaudhuri equation and is a core element of the Hawking-Penrose singularity theorems in GR. In this way, the Raychaudhuri equation essentially implicates the idea that a singularity can simply be a byproduct of symmetries present in the matter distribution under consideration. The equation is a geometrical relation which governs the dynamics of mean separation between a congruence of curves. The equation and it's generalizations have found significant application in gravitational physics, for instance, validation of singularity theorems, gravitational lensing, cracking of self-gravitating compact objects, derivation of the equations of thermodynamics of spacetime.\\

In a gravitational system, the proper acceleration represented by $d^2 x^{\mu}/d\tau^2$ is an observer dependent quantity and not covariant, and so may vanish for some observers and not for others. For this reason, one must take the acceleration as
\begin{equation}\label{geo_eq}
a^{\mu} = u_{\alpha ;\beta}u^{\beta} = \frac{D^2 x^{\mu}}{D\tau^2} = \frac{d^2 x^{\mu}}{d\tau^2} +\Gamma^{\mu}_{\nu\sigma}u^{\nu}u^{\sigma}\,,
\end{equation}
for it to be covariant in nature, where $D$ represents a covariant derivative and $u^{\nu}=d x^{\mu}/d \tau$ represents the four-velocity of a particle in the system. For a congruence of curves $x^{\mu}(\tau; \lambda)$, where $\lambda$ parametrizes the paths that satisfy Eq.(\ref{geo_eq}), the four-velocity $u^{\mu}(\tau, \lambda)$ can be interpreted as one of the tangent fields together with $n^{\mu}=d x^{\mu}/d \lambda (\tau, \lambda)$ \cite{Harko:2012ve}.\\

The Raychaudhuri equation is derived by considering points on infinitely close geodesics corresponding to the parametrization values $\lambda$ and $\lambda + \delta \lambda$. Within this determination, the trace expansion, rotation tensor and shear are respectively defined as
\begin{align}
    \theta &= \udt{u}{\mu}{;\mu}\,,\\
    \omega_{\alpha\beta} &= \nabla_{[\mu} u_{\nu]} - \dot{u}_{[\mu}u_{\nu]}\,,\\
    \sigma_{\alpha\beta} &= \nabla_{(\alpha} u_{\beta)} - \frac{1}{3}\theta h_{\alpha\beta} - \dot{u}_{(\alpha} u_{\nu)}\,,
\end{align}
where the projection tensor is defined as $h_{\mu\nu} = g_{\mu\nu} - u_{\mu} u_{\nu}$. In this scenario, the expansion scalar is defined as the fractional rate of change of volume of a matter distribution measured by a comoving observer defined as the fractional rate of change of volume of a matter distribution measured by a comoving observer. If this derivative is negative along some worldline then the matter distribution must be collapsing. The shear and the rotation tensor measures the distortion and rotation of an initially spherical matter distribution. Now, by considering the congruence of timelike geodesics leads to the Raychaudhuri equation which reads as \cite{1955PhRv...98.1123R,misner1973gravitation}
\begin{equation}\label{Raych-time}
 \frac{d\theta}{d\tau} = -\frac{1}{3}\theta^2 + \nabla_{\alpha}a^{\alpha} - \sigma_{\alpha\beta}\sigma^{\alpha\beta} + \omega_{\alpha\beta}\omega^{\alpha\beta} - \lc{R}_{\alpha\beta}u^\alpha u^\beta\,,
 \end{equation}
where $\lc{R}_{\alpha\beta}$ is the Ricci tensor (determined with the Levi-Civita connection). It is important to note that the appearance of the Levi-Civita connection in the Raychaudhuri equation does not emerge from it being the connection for GR but due to the way that the equation is derived. For this reason, the Ricci tensor continues to be derived using the Levi-Civita connection within TG. Moreover, the general theorems regarding Riemann manifolds continue to hold and so the appearance of the standard gravity Ricci tensor does not cause any consistency conflicts within this regime. \\

The evolution equation for the expansion of a congruence of null geodesics defined by a null vector field $k^{\mu}$ ($k^{\mu}k_{\mu}=0$) has a similar form as the Raychaudhuri equation in Eq.(\ref{Raych-time}), but with a factor $1/2$ rather than $1/3$, and $-R_{\mu\nu}k^{\mu}k^{\nu}$ instead of $-R_{\mu\nu}u^{\mu}u^{\nu}$ as the last term. Thus,
it reads as
\begin{equation}\label{Raych-null}
\frac{d\theta}{d\tau}= -\frac{1}{2}\,\theta^2 - \sigma_{\mu\nu}\sigma^{\mu\nu} + \omega_{\mu\nu}\omega^{\mu\nu} - R_{\mu\nu}k^\mu k^\nu \,,
\end{equation}
where the kinematical quantities $\theta\,$, $\sigma^{\mu\nu}$ and $\omega_{\mu\nu}$ are now clearly associated with the congruence of null geodesics. An important point to be emphasized is that Eqs.~(\ref{Raych-time}) and~(\ref{Raych-null}) are purely geometric statements, and as such they make no reference to any theory of gravitation in that they are general results for Riemann manifolds, and thus would be applicable to all other connection constructs within this framework.

\section{\label{TG_background} Teleparallel Gravity and its Extensions} 
TG is a novel reformulation of gravitation in that the curvature associated with the manifestation of gravity is exchanged with torsion \cite{aldrovandi1995introduction,Cai:2015emx,Krssak:2018ywd} through the replacement of the Levi-Civita connection, $\mathring{\Gamma}{}^{\sigma}{}_{\mu\nu}$, with the so-called Weitzenb\"{o}ck connection expressed through \cite{Weitzenbock1923,Aldrovandi:2013wha}
\begin{equation}\label{eq:weitzenbockdef}
\udt{\Gamma}{\sigma}{\mu\nu} := \dut{e}{a}{\sigma}\partial_\mu \udt{e}{a}{\nu} + \dut{e}{a}{\sigma}\udt{\omega}{a}{b\mu}\udt{e}{b}{\nu}\,,
\end{equation}
where $\udt{e}{a}{\rho}$ is the tetrad field ($\dut{e}{a}{\mu}$ being the transpose), and $\udt{\omega}{a}{b\mu}$ the spin connection (over-circles are used on all quantities calculated with the Levi-Civita connection). In fact, there exists a trinity of possible ways to express gravity through geometry with the third being based on nonmetricity rather than curvature or torsion \cite{BeltranJimenez:2019tjy}. In all these cases there exists a limit in which these formulations limit to GR in that they produce the identical dynamical equations (despite having different actions due to the appearance of a boundary term).

The Weitzenb\"{o}ck connection is the most general linear affine connection that is both curvatureless and satisfies the metricity condition \cite{aldrovandi1995introduction}. The exact expression of the Weitzenb\"{o}ck connection depends on the tetrad, $\udt{e}{a}{\rho}$, and the inertial spin connection, $\udt{\omega}{a}{b\mu}$. The tetrad acts as a soldering agent between the general manifold and the tangent (inertial) space which are represented by Greek and Latin indices respectively. The spin connection sustains the invariance of the field equations under local Lorentz transformations (LLTs) \cite{Li:2010cg}. The spin connection is a crucial ingredient which must appear in the field equations due to use of tetrads since they have one inertial index, rather than being an extra degree of freedom of the theory. Together the tetrad and spin connection describe spacetime in TG in the same way that the metric tensor does so in GR, and are thus the fundamental dynamical object of the theory.

Considering the full breadth of possible LLTs (boosts and rotations), $\udt{\Lambda}{a}{b}$, the tetrads are transformed on the tangent space by
\begin{equation}
\udt{e'}{a}{\mu}=\udt{\Lambda}{a}{b}\udt{e}{b}{\mu}\,,
\end{equation}
whereas the spin connection transformed as
\cite{Krssak:2015oua}
\begin{equation}
\udt{\omega}{a}{b\mu} = \udt{\Lambda}{a}{c}\partial_{\mu}\dut{\Lambda}{b}{c}\,,
\end{equation}
which together preserve the LLTs of the theory as a whole. On the other hand, there also exist so-called good tetrads which organically produce vanishing spin connection components \cite{Tamanini:2012hg,Bahamonde:2017wwk}. However, given the LLT of the theory, all consistent tetrad and spin connection pairs will be dynamically equivalent in terms of the field equations they produce.

In TG, the tetrad embodies the effect of gravity in a similar way as the metric tensor expresses geometric deformation in curvature-based theories of gravity \cite{Cai:2015emx,Krssak:2018ywd}. For consistency, the tetrads observe the relations \cite{Aldrovandi:2013wha}
\begin{align}
\udt{e}{a}{\mu}\dut{e}{b}{\mu} = \delta^a_b\,, \quad
\udt{e}{a}{\mu}\dut{e}{a}{\nu} = \delta^{\nu}_{\mu}\,,
\end{align}
which form the orthogonality conditions of the tetrad fields. Since the effect of tetrads is to connect the general manifold and its Minkowski space, this can be used to transform between these spaces. One example of this is with the Minkowski metric which transforms as
\begin{align} \label{Min_trans_Gen}
g_{\mu\nu} = \udt{e}{a}{\mu}\udt{e}{b}{\nu}\eta_{ab}\,,\quad
\eta_{ab} = \dut{e}{a}{\mu}\dut{e}{b}{\nu}g_{\mu\nu}\,.
\end{align}
The position dependence of these relations is being suppressed for brevity's sake.

TG is fundamentally distinct from curvature-based descriptions of gravity in that the exchange of the Levi-Civita with its analog Weitzenb\"{o}ck connection means that all measures of curvature (such as the Riemann tensor and Ricci scalar) will organically vanish for the torsional case \cite{aldrovandi1995introduction}. Thus, TG requires a wholly different formulation on which to quantify the effect of gravity. In this setting, torsion is measured through the torsion tensor which is represented as an antisymmetric operation \cite{Bahamonde:2017wwk}
\begin{equation}
\udt{T}{\sigma}{\mu\nu} :=  2\udt{\Gamma}{\sigma}{[\mu\nu]}\,,
\end{equation}
which also serves as the field strength of gravitation (square brackets represent the anti-symmetric operator $A_{[\mu\nu]}=\frac{1}{2}\left(A_{\mu\nu}-A_{\nu\mu}\right)$). The torsion tensor transforms covariantly under both diffeomorphisms and LLTs, and observes the anti-symmetry $\udt{T}{\sigma}{\mu\nu} = -\udt{T}{\sigma}{\nu\mu}$.

The torsion tensor, analogous to the Riemann tensor, is a measure of torsion for a gravitational field. However, other important and useful quantities exist in TG. The contorsion tensor is one such quantity measure; this is determined as the difference between the Levi-Civita and Weitzenb\"{o}ck connections \cite{Cai:2015emx,RevModPhys.48.393}
\begin{equation}\label{Contorsion_def}
\udt{K}{\sigma}{\mu\nu} := \Gamma^{\sigma}{}_{\mu\nu} - \mathring{\Gamma}{}^{\sigma}{}_{\mu\nu} = \frac{1}{2}\left(\dudt{T}{\mu}{\sigma}{\nu}+\dudt{T}{\nu}{\sigma}{\mu}-\udt{T}{\sigma}{\mu\nu}\right)\,.
\end{equation}
The contorsion tensor is crucial to relating TG with its curvature-based analogues. Along a similar vein, the so-called superpotential is another TG tensor of central importance which is defined as \cite{Aldrovandi:2013wha}
\begin{equation}\label{Super_def}
\dut{S}{a}{\mu\nu} := \frac{1}{2}\left(\udt{K}{\mu\nu}{a} - \dut{h}{a}{\nu}\udt{T}{\alpha\mu}{\alpha} + \dut{h}{a}{\mu}\udt{T}{\alpha\nu}{\alpha}\right)\,.
\end{equation}
The superpotential may play a critical role in reformulating TG as a gauge current for a gravitational energy-momentum tensor \cite{Aldrovandi:2004db,Koivisto:2019jra}. Also, the superpotential observes the anti-symmetry $\dut{S}{a}{\mu\nu} = -\dut{S}{a}{\nu\mu}$.

Contracting the torsion and the superpotetial tensors produces the torsion scalar through \cite{Krssak:2018ywd}
\begin{equation}\label{torsin_ten_def}
T := \dut{S}{a}{\mu\nu}\udt{T}{a}{\mu\nu}\,,
\end{equation}
which is determined solely by the Weitzenb\"{o}ck connection, in the same way that the Ricci scalar is determined by the Levi-Civita connection. Interestingly, it coincidentally turns out that the Ricci and torsion scalars are equal up to a total divergence term \cite{Bahamonde:2017ifa,Bahamonde:2015zma}
\begin{equation}\label{TEGR_L}
R=\mathring{R} +T-\frac{2}{e}\partial_{\mu}\left(e\udut{T}{\sigma}{\sigma}{\mu}\right)=0\,,
\end{equation}
where $R$ is the Ricci scalar as calculated with the Weitzenb\"{o}ck connection which naturally vanishes. Thus, it follows that
\begin{equation}\label{Ricc_ten_iden}
\mathring{R} = -T + \frac{2}{e}\partial_{\mu}\left(e\udut{T}{\sigma}{\sigma}{\mu}\right) := -T + B\,,
\end{equation}
where $\mathring{R}$ is the Ricci scalar as determined using the Levi-Civita connection, and $e$ is the determinant of the tetrad field, $e = \det\left(\udt{e}{a}{\mu}\right) = \sqrt{-g}$. Here, $B$ embodies the boundary term. This equivalency alone guarantees that the variation of the torsion and Ricci scalars produce the same dynamical equations. Also, this means that the second- and fourth-order contributions to the Ricci scalar can be decoupled from each other in TG, which may have important consequences for producing a more natural generalization of $f(R)$ gravity \cite{Sotiriou:2008rp,Capozziello:2011et,Clifton:2011jh}.

Another natural consequence of this equivalency is that the TEGR action can be defined directly as \cite{Bahamonde:2017wwk}
\begin{equation}
\mathcal{S}_{\rm TEGR} = -\dfrac{1}{2\kappa^2} \int d^4x \: e T + \int d^4x \: e \mathcal{L}_{\rm m}\,,
\end{equation}
where $\kappa^2 = 8\pi G$, and $\mathcal{L}_{\rm m}$ represents the Lagrangian for matter. Despite being described through the tetrad and spin connection, TEGR will produce identical dynamical equations as GR, namely
\begin{align}\label{TEGR_FEs}
    \mathring{G}_{\mu\nu} &\equiv e^{-1}\udt{e}{a}{\mu}g_{\nu\rho}\partial_\sigma(e \dut{S}{a}{\rho\sigma})-\dudt{S}{b}{\sigma}{\nu}\udt{T}{b}{\sigma\mu} \nonumber\\
    &+\frac{1}{4}T g_{\mu\nu}-\udt{e}{a}{\mu} \udt{\omega}{b}{a\sigma}\dut{S}{b\nu}{\sigma} = \kappa^2 \Psi_{\mu\nu}\,,
\end{align}
where $\Psi_{\mu\nu}$ is the energy-momentum tensor \cite{misner1973gravitation} given by $\dut{\Psi}{a}{\mu} := \delta \mathcal{L}_{\rm m}/\delta \udt{e}{a}{\mu}$, and $\mathring{G}_{\mu\nu}$ is the Einstein tensor calculated with the Levi-Civita connection. Also, it is important to point out that while the Weitzenb\"{o}ck connection is used in th gravity sector, the Levi-Civita connection continues to feature in the coupling prescription of matter \cite{Aldrovandi:2013wha,Krssak:2018ywd}.

Taking the same path of modification as in $f(\mathring{R})$ gravity, the TEGR Lagrangian can be generalized to $f(T)$ gravity \cite{Ferraro:2006jd,Ferraro:2008ey,Bengochea:2008gz,Linder:2010py,Chen:2010va}, namely $\mathcal{L}_{f(T)}=e f(T)$. By taking a variation with the tetrad, this results in field equations
\begin{align}
    e^{-1} &\partial_{\nu}\left(e\dut{e}{a}{\rho}\dut{S}{\rho}{\mu\nu}\right)f_T - \dut{e}{a}{\lambda} \udt{T}{\rho}{\nu\lambda}\dut{S}{\rho}{\nu\mu} f_T + \frac{1}{4}\dut{e}{a}{\mu}f(T) \nonumber\\
    & + \dut{e}{a}{\rho}\dut{S}{\rho}{\mu\nu}f_{TT}\partial_{\nu}\left(T\right)\nonumber\\
    & + \dut{e}{b}{\lambda}\udt{\omega}{b}{a\nu}\dut{S}{\lambda}{\nu\mu}f_T = \kappa^2 \dut{e}{a}{\rho} \dut{\Psi}{\rho}{\mu}\,,\label{f_T_FEs}
\end{align}
which can be contracted with the tetrad or metric tensor depending on whether the index in question is a tangent space or general manifold label. These field equations are generically second-order in nature \cite{Cai:2015emx}, as well as a number of other similarities to GR such as their associated gravitational waves exhibiting identical polarizations \cite{Farrugia:2018gyz,Abedi:2017jqx}. However, to incorporate a framework on which to compare results with their $f(\mathring{R})$ gravity analog, we must consider $f(T,B)$ gravity \cite{Bahamonde:2015zma,Capozziello:2018qcp,Bahamonde:2016grb,Paliathanasis:2017flf,Farrugia:2018gyz,Bahamonde:2016cul,Bahamonde:2016cul,Wright:2016ayu,Farrugia:2020fcu} in which the decoupled second- and fourth-order contributions appear in the torsion scalar and boundary term respectively. In these cases, the limit to $f(\mathring{R})$ gravity occurs for the consideration $f(T,B)=f(-T+B)=f(\mathring{R})$ gravity. Another interesting avenue on which to construct modified teleparallel theories of gravity is to consider nonminimal couplings with matter \cite{Farrugia:2016pjh,Momeni:2014jja,Saez-Gomez:2016wxb,Nassur:2015zba,Pace:2017dpu,Pace:2017aon,Harko:2014aja,Bahamonde:2017ifa}. Given the organically lower-order nature of the torsion scalar means that such modification to gravity may produce novel observational consequences. 

In what follows, we choose frames where the spin connection is allowed to vanish. Since a frame always exists where this is possible, we do not overly limit the applicability of this work. Also, we take units where $\kappa^2 = 1$.\\

We now focus on the $f(T)$ field equations written in Eq.(\ref{f_T_FEs}) which can be written using only general manifold indices to give
\begin{align}
    e^{-1} &g_{\mu\sigma} \udt{e}{a}{\nu} \partial_{\gamma}\left(e\dut{e}{a}{\rho}\dut{S}{\rho}{\sigma\gamma}\right)f_T - \udt{T}{\rho}{\gamma\nu}\dut{S}{\rho}{\gamma\sigma}g_{\sigma\mu}f_T \nonumber\\
    &+ \frac{1}{4}g_{\mu\nu} f(T) + g_{\sigma\mu}\dut{S}{\nu}{\sigma\gamma}f_{TT}\partial_{\gamma}T = \Psi_{\mu\nu}\,,
\end{align}
which follows by raising the inertial index with an inverse tetrad. This gives a trace equation
\begin{align}
    e^{-1}&\udt{e}{a}{\sigma}\partial_{\gamma}\left(e\dut{e}{a}{\rho}\dut{S}{\rho}{\sigma\gamma}\right)f_T - Tf_T + f(T) \nonumber\\
    & + \dut{S}{\sigma}{\sigma\gamma}f_{TT}\partial_{\gamma}T = \Psi\,.
\end{align}
Using the TEGR field equations in Eq.(\ref{TEGR_FEs}), we can also write these field equations down using the standard Einstein tensor as
\begin{align}
    \mathring{G}_{\mu\nu}f_T + \frac{1}{4}g_{\mu\nu}\left(f(T) - Tf_T\right) + g_{\sigma\mu} \dut{S}{\nu}{\sigma\gamma} f_{TT}\partial_{\gamma}T = \Psi_{\mu\nu}\,,\nonumber\\
\end{align}
which has an interesting trace equation
\begin{equation}
    f(T)-\left(\mathring{R}+T\right)f_T + \dut{S}{\sigma}{\sigma\gamma}f_{TT}\partial_{\gamma}T = \Psi\,,
\end{equation}
where the term in parenthesis turns out to be $\mathring{R}+T=B$ using Eq.(\ref{Ricc_ten_iden}).

Given the Einstein tensor definition, namely $\mathring{G}_{\mu\nu}:=\mathring{R}_{\mu\nu} - \frac{1}{2} g_{\mu\nu}\mathring{R}$, the Ricci tensor dependency on the $f(T)$ Lagrangian can be expressed as
\begin{align}\label{Ricci_ten_comp}
    \mathring{R}_{\mu\nu} &= \frac{1}{2}g_{\mu\nu}\left(-T+B\right)\nonumber\\
    & + \frac{1}{f_T}\Bigg[\Psi_{\mu\nu} - \frac{1}{4}g_{\mu\nu} \left(f(T) - Tf_T\right)\nonumber\\
    & - g_{\sigma\mu}\dut{S}{\nu}{\sigma\gamma} f_{TT} \partial_{\gamma} T\Bigg]\,,
\end{align}
which can now be used with the Raychaudhuri equations in Eq.(\ref{Raych-null}) to determine the effect of $f(T)$ gravity on the congruence of null geodesics.

\section{\label{collapse} A Collapsing Spherical Star in \texorpdfstring{$f(T)$}{ft} gravity}
We consider a spatially homogeneous collapsing stellar distribution whose interior is described by the metric
\begin{equation}\label{stellar_collap}
    ds^2 = dt^2 - a(t)^{2} \left[dx^2 + dy^2 + dz^2\right]\,,
\end{equation} 
where $a(t)$ is the physical radius of the collapsing system. The tetrad choice $\udt{e}{a}{\mu}={\rm diag}(1,a(t),a(t),a(t))$ is compatible with a vanishing spin connection \cite{Krssak:2015oua}. On the other hand, we take the energy-momentum contribution to be that of a perfect fluid described by
\begin{eqnarray}\label{emt}
    \Psi_{\mu\nu}=(\rho+p)u_\mu u_\nu-p\,g_{\mu\nu} \;\; \text{with} \;\; u_\mu=(1,0,0,0)\,,
\end{eqnarray}
where $k_{\mu}=\left(1,a,0,0\right)$ for the null vector field in Eq.(\ref{Raych-null}), $p$ is the fluid pressure and $\rho$ its energy density. Thus, the Raychaudhuri equation can be written as
\begin{equation} \label{rcfinal}
\frac{d\theta}{d\tau} = -\frac{\theta^2}{3} - \frac{1}{2f_T}(\rho+3p+f-Tf_T-72H^2\dot{H}f_{TT})\,,
\end{equation}
which describes the congruence of neighbouring particle geodesics, and where $H=\dot{a}/a$, and dots refer to derivatives with respect to time, $t$. The field equations can then be written as
\begin{align}\label{fe}
\rho &= 6H^2 f_{T} + \frac{f}{2}\,, \\
(\rho + p) &= 24H^2 \dot{H} f_{TT} - 2\dot{H}f_{T}\,.
\end{align}
Using Eqs. (\ref{fe}) in Eq. (\ref{rcfinal}), and putting $\theta = 3\frac{\dot{a}}{a}$, we write
\begin{equation}\label{congruence}
\frac{d\theta}{d\tau} = 3 \dot{H} = -\frac{3(\rho + p)}{2(2Tf_{TT} + f_{T})}\,.
\end{equation}
 
Eq. (\ref{congruence}) governs the evolution of the time-like congruence, depending on the positivity or negativity of $\frac{d\theta}{d\tau}$. A negative $\frac{d\theta}{d\tau}$ indicates a throughout collapsing system until $\theta$ reaches $-\infty$, indicating a zero proper volume singularity. However, if $\frac{d\theta}{d\tau}$ changes signature to positive over the course of it's evolution then a collapse of the congruence is halted and the geodesics start to move away from each other. Therefore, the formation of a zero proper volume singularity may be avoided. The onus of avoiding a singularity therefore lies on the behavior of the RHS of Eq. (\ref{congruence}).  \\

If we assume that both the energy density $\rho$ and the isotropic pressure $p$ are positive in nature, the evolution of the congruence and the predictibility of the collapse depends entirely on the nature of $(2Tf_{TT} + f_{T})$, i.e., explicitly dependent on the choice of $f(T)$ one makes. If $(2Tf_{TT} + f_{T})>0$, the congruence is collapsing and if $(2Tf_{TT} + f_{T})<0$, the congruence is expanding. Therefore the predictibility of a collapsing stellar distribution in $f(T)$ theories depends on the choice of $f(T)$ as a congruence of time-like geodesics would suggest. Using the definition of the torsion scalar in Eq.(\ref{torsin_ten_def}), it follows that $T = -6H^2$, throughout which one can write the RHS of Eq. (\ref{congruence}) in terms of $f = f(H)$ which simplifies this equation into
\begin{equation}\label{congruence1}
\frac{d\theta}{d\tau} = 3 \dot{H} = \frac{18(\rho + p)}{f_{HH}}\,.
\end{equation}
Thus, the evolution of the congruence depends on the nature of $f_{HH}$. If $f_{H}$ is a decreasing function of $H$, the geodesics are imploding towards one another until a singularity is formed. If during the evolution, $f_{H}$ becomes an increasing function of $H$, the collapse halts and the geodesics start to move apart from each other. \\

Positivity of both the energy density and pressure implies that $(\rho + p) > 0$, which is the usual Null Energy Condition (NEC) in the context of GR. In the context of an $f(T)$ theory the NEC can be written from Eq. (\ref{Raych-null}) as
\begin{equation}\label{nec0}
\lc{R}_{\mu\nu}k^\mu k^\nu \geq 0\,.
\end{equation}
The NEC is a general result of Riemann manifolds rather than GR which is why it continues to be expressed in terms of th standard gravity Ricci tensor. It is for this reason that it retains its dependence on the Levi-Civita connection rather than the Weitzen\"{o}ck connection. This form of the NEC statement is essentially a coordinate-invariant way for an unfixed geometrical theory of gravitation. For a general $f(T)$ gravity this can be written as 
\begin{equation}\label{nec1}
\frac{1}{f_T}(\rho+p-24H^2\dot{H}f_{TT})\geq 0 \,,
\end{equation}
where the positivity of both the energy density and pressure helps one to ensure that the NEC is also satisfied throughout the evolution. In the following works, we plot the LHS of the NEC in a simple example to discuss the evolution of the matter distribution for a simple collapsing exact solution. It is quite natural in a study of gravitational collapse to plot the NEC as a function of time as was first done in the vintage paper of Kolassis, Santos and Tsoubelis \cite{Kolassis_1988}. The idea is to write the LHS of NEC as a function of time for different collapsing shells labelled by different values of radial distance r. In case of a spatially homogeneous meric as in our case, this makes the LHS of NEC a function of time (See also \cite{2014PhRvD..90h4011G}). \\

Since we are completely avoiding the rigorous avenue of finding an exact solution, more analysis relies heavily on the amount of information that can be extracted from the Raychaudhuri Eq. (\ref{congruence}). There are a few different ways of analysing further, for instance, one can choose a certain behavior of $\frac{d\theta}{d\tau}$ and solve the resulting equation. As the simplest possible example, let us assume that $(2Tf_{TT} + f_{T}) = \delta$, where $\delta$ is a constant. This can be solved straightaway to write
\begin{equation}\label{ft1}
f(T) = \delta T + 2 C_{1} T^{1/2} + C_{2}. 
\end{equation} 
Thus, depending on a positive or a negative $\delta$, a time-like congruence under the scope of an $f(T)$ theory (Eq. (\ref{ft1})) avoids a formation of zero proper volume singularity or not. \\

The second method is to choose a particular viable $f(T)$ model from literature, and explore what constraints te Raychaudhuri equation enforces upon the model parameters. To this end, we choose two popular models of $f(T)$ gravity. For a power law $f(T)$ theory \cite{Bengochea:2008gz,Nesseris:2013jea}
\begin{equation}\label{ft2}
f(T) = T + \alpha (-T)^{n}.
\end{equation}

From Eqs. (\ref{congruence}) and (\ref{ft2}), we find the relation governing collapsing or expanding nature of the geodesic congruence. An initially collapsing congruence remains collapsing if 
\begin{equation}
1 - [2\alpha n(n-1) + \alpha n](-T)^{n-1} > 0\,,
\end{equation} 
and it changes nature from collapsing to expanding if
\begin{equation}
1 - [2\alpha n(n-1) + \alpha n](-T)^{n-1} < 0\,.
\end{equation} 
Similarly, for an exponential $f(T)$ model, given by \cite{Linder:2010py}
\begin{equation}\label{ft3}
f(T) = T + \alpha T_{0} \Bigg[1 - e^{-p\sqrt{\frac{T}{T_{0}}}}\Bigg], 
\end{equation}
we find the relation governing collapsing or expanding nature of the geodesic congruence using Eq. (\ref{congruence}). If
\begin{equation}
1 - \frac{\alpha p^2}{2} e^{-p\sqrt{\frac{T}{T_{0}}}} > 0,
\end{equation} 
the congruence is collapsing towards a formation of singularity. If during the course of it's evolution, the LHS of the above equation changes signature and satisfies 
\begin{equation}
1 - \frac{\alpha p^2}{2} e^{-p\sqrt{\frac{T}{T_{0}}}} < 0,
\end{equation} 
the congruence becomes expandng, completely avoiding any formation of zero proper volume. \\

To elaborate a little more rigorously, we also try to explore from a different point of view and propose that an initially collapsing congruence can be described by a spatially homogeneous metric with a scale factor
\begin{equation}\label{paramet}
a(t) = \delta_{1} e^{\alpha f(t)} + \delta_{2} e^{-\alpha f(t)}\,.
\end{equation}

The form in Eq. (\ref{paramet}) defines a kind of parametrization and we argue that this form can give a general evolution of all the possible outcomes of an initially collapsing stellar distribution for different form of the function $f(t)$ and for different values of the parameters $\delta_{1}$, $\delta_{2}$ and $\alpha$. $f(t)$ is a continuous and differentiable function of time, which can be exactly determined by writing an exact solution, if possible, from the modified field equations. However, finding an exact solution of the field equations may be extremely non-trivial and not part of the purpose of this work. We mean to comment on the restrictions one must impose on the function $f(t)$ and the parameters such that different time evolution paths are described. Using Eq. (\ref{paramet}) in Eq. (\ref{congruence}) we can write
\begin{equation}\label{paramet2}
\frac{d\theta}{d\tau} = 3 \Bigg[\alpha \ddot{f} \Bigg(1 - \frac{4\delta_{1}\delta_{2}}{a^{2}}\Bigg)^{1/2} + \frac{4 \delta_{1}\delta_{2} \alpha^{2}\dot{f}^{2}}{a^2}\Bigg]\,.
\end{equation}

\begin{figure}
\begin{center}
\includegraphics[width=0.40\textwidth]{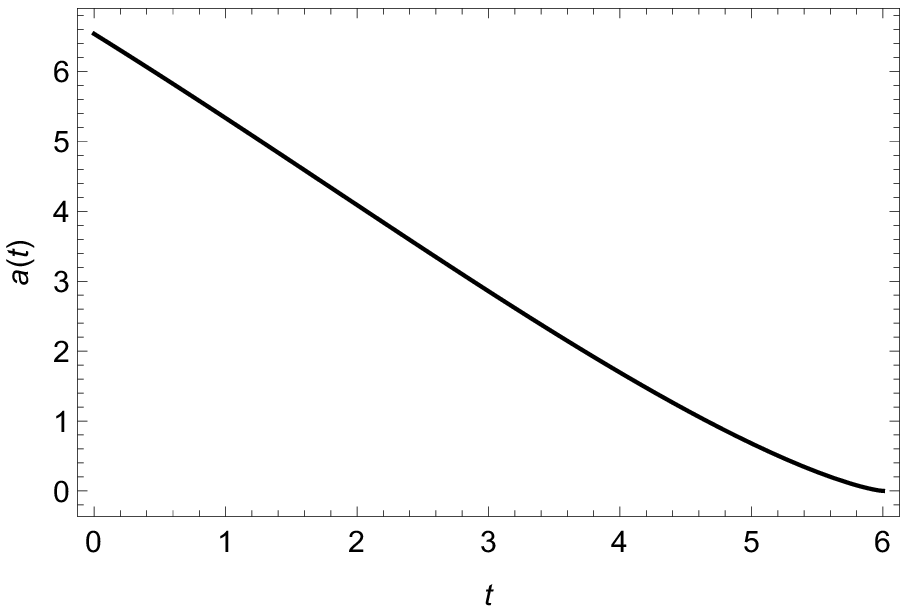}
\includegraphics[width=0.41\textwidth]{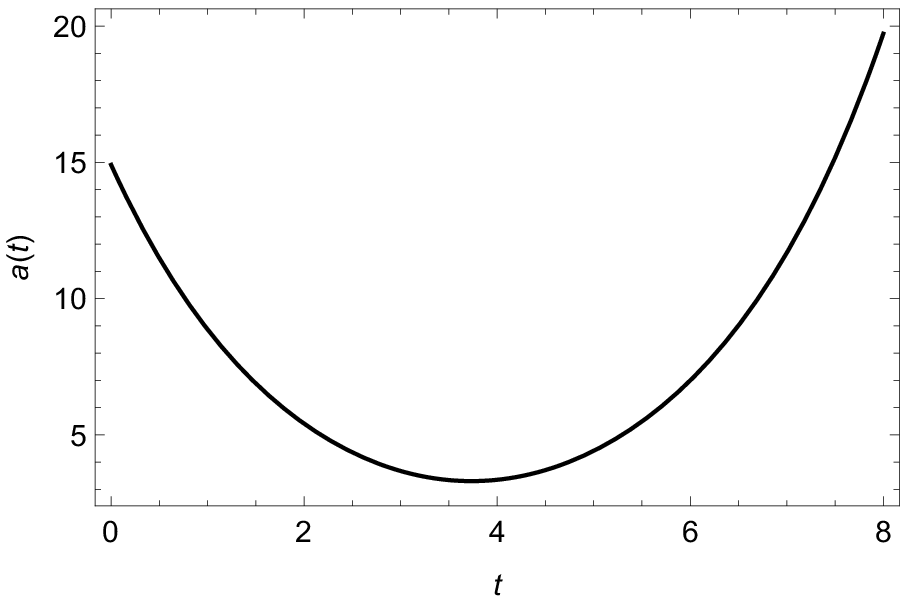}
\caption{Evolution of the radius of two-sphere as given in Eq. (\ref{exactminimal}) : (i) Top curve $\alpha = 0.01$, $C = 1$ and $a_0 = 100$ ; (ii) Bottom curve $\alpha = 0.01$, $C = 1$ and $a_0 = -100$.}
\label{Fig1}
\end{center}
\end{figure}

Any chance of a bounce of the initially collapsing star depends on the existence of zero-s of the RHS of Eq. (\ref{paramet2}). There may be one or more of such zeros where the evolution changes nature to expanding from collapsing and vice-versa. Depending on the number of zeros, an initially collapsing star can continue to collapse, it can bounce after a finite time, or it can suffer a multiple of collapse-and-bounce segments and eventually become oscillatory in nature (See Fig. \ref{Fig1} for graphical representations for different choices of $f(t)$). Putting $\frac{d\theta}{d\tau} = 0$ in Eq. (\ref{paramet2}) therefore yields a `critical condition' written as
\begin{equation}\label{paramet3}
\ddot{f} = -\dot{f}^{2} \frac{4\alpha \delta_{1} \delta_{2}}{(a^{2} - 4\delta_{1} \delta_{2})^{1/2}}\,.
\end{equation}

We can summarize the possible outcomes of the collapse from the critical condition Eq. (\ref{paramet3}) as follows
\begin{enumerate}
\item {The value of $\ddot{f}$ crossing the limit of $-\dot{f}^{2} \frac{4\alpha \delta_{1} \delta_{2}}{(a^{2} - 4\delta_{1} \delta_{2})^{1/2}}$ is a signature of change from collapse to a probable expansion/dispersal or vice versa for a collapsing star parametrized by Eq. (\ref{paramet}).}
\item {Since $\dot{f}$ and $\ddot{f}$ are real functions of time, the RHS of Eq. (\ref{paramet3}) must also be real. If any one of $\delta_{1}$ or $\delta_{2}$ is negative, $(a^{2} - 4\delta_{1} \delta_{2})^{1/2}$ is always real and $a(t)$ can evolve on to a zero proper volume, forming a zero proper volume singularity.}
\item{However, if $\delta_{1}$ and $\delta_{2}$ are both positive, RHS of Eq. (\ref{paramet3}) is real if and only if $(a^{2} - 4\delta_{1} \delta_{2}) > 0$. This predicts a different outcome of the collapse and notes it's sensitivity on the choice of initial parameters. In such a case, the minimum allowed value of $a(t)$ is $a_{critical} = 2 \delta_{1}^{1/2} \delta_{2}^{1/2}$. Beyond this critical radius no more shrinking of the congruence is allowed and a bounce/dispersal must take place. Thus, the effective modification of GR due to the non-conservation of energy-momentum distribution opens up more possibilities regarding the end-state of gravitational collapse as compared to standard GR.}
\end{enumerate}

As an example, we present a particular exact solution of the field equations in Eq. (\ref{fe}) for $f(T) = T + \alpha (-T)^2$. This is a special case of a more general theory given by $f(T) = T + \alpha (-T)^n$. For a perfect fluid given by $p = \omega \rho$ describing the collapsing fluid distribution, manipulating the field equations we write
\begin{eqnarray}\nonumber
&&\dot{H} \left\lbrace 48 \alpha H^2 - 2 + 24 \alpha H^2 \right\rbrace - 6(1+\omega)H^{2} \left(1-12\alpha H^2 \right) \\&&
+ 3H^{2}(1+\omega) -18(1+\omega)\alpha H^4 = 0\,. 
\end{eqnarray}

During the final phases of the collapse, $\dot{a} \rightarrow \infty$ and $a \rightarrow 0$, therefore $H >> \frac{1}{\alpha}$. In this limit the above equation can be written as
\begin{equation}
\dot{H} = -\frac{3}{4} (1+\omega) H^2 + \frac{1}{24\alpha} (1+\omega)\,.
\end{equation}
A first integral of the above equation can be calculated as
\begin{equation}
\dot{a} = - \left\lbrace a_{0} a^{-m} + \frac{(1+\omega) a^2}{24\alpha (m+2)} \right\rbrace ^{\frac{1}{2}}\,.
\end{equation} 
$a_0$ is a parameter related to the initial value of $\dot{a}$ and $m = \frac{3}{2}(1+\omega) - 2$. As a simple example we solve the first integral equation for $\omega = 1$, which implies $m = 1$. For this we write the evolution of the radius of two-sphere as
\begin{equation}\label{exactminimal}
a(t) = \frac{e^{\frac{6C - t}{6\sqrt{\alpha}}}}{2^{2/3}} \left[e^{\frac{3C}{\sqrt{\alpha}}} - 36 a_{0}\alpha e^{\frac{t}{2\sqrt{\alpha}}} \right]^{\frac{2}{3}}\,.
\end{equation}

In Fig. \ref{Fig1} we plot the evolution as a function of time. The graph on top of the figure shows a plot for $a_0 > 0$. It is clear to note that the collapsing fluid reaches a zero proper volume at a finite future. The time of formation of this zero proper volume singularity may vary depending on the choice of the functional form of the theory, i.e., $\alpha$, however, the qualitatiove behavior remains the same. In the graph below, the evolution is shown for $a_0 < 0$. It is clear that there is no formation of zero proper volume singularity in this case, as the collapsing fluid bounces indefinitely after reaching a minimum cutoff volume. The parameter $a_0$ is a critical parameter of the system whose signature determines the fate of the collapsing system. Using the equation for the NEC, as in Eq. (\ref{nec1}), we check if the collaping fluid satisfies the NEC. This essentially ensures a positive energy density and that the speed of energy flow of matter is less than the speed of light. 

\begin{figure}
\begin{center}
\includegraphics[width=0.40\textwidth]{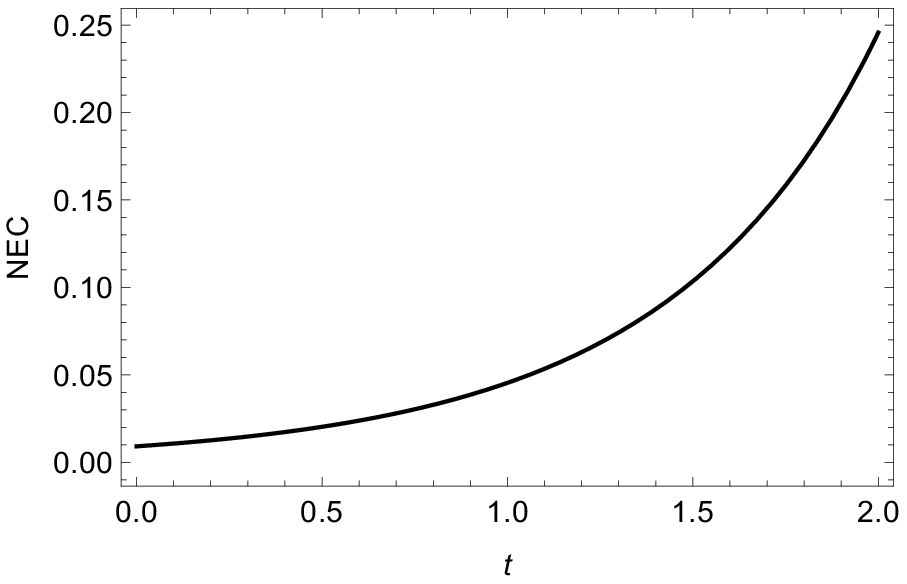}
\includegraphics[width=0.41\textwidth]{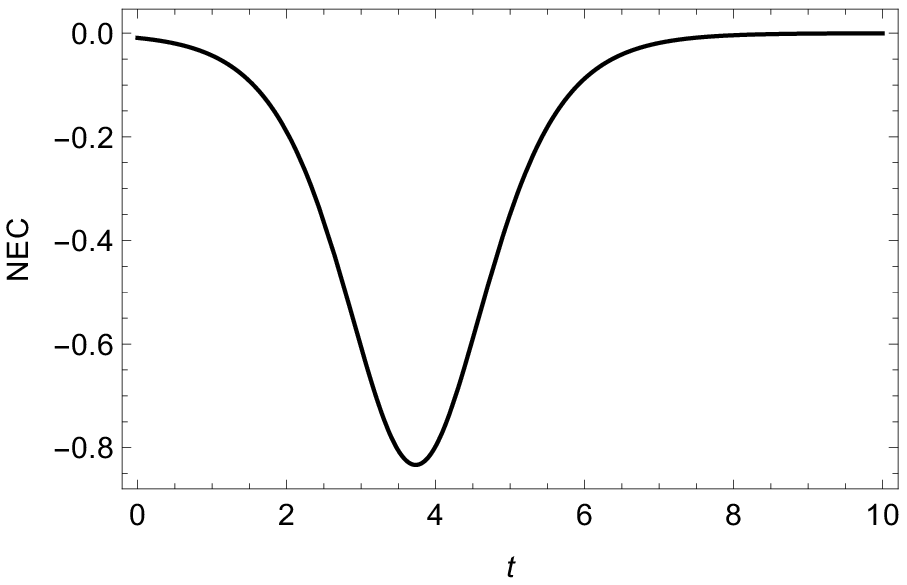}
\caption{Evolution of the NEC as given in Eq. (\ref{nec1}) : (i) Top curve shows the evolution for collapse, i.e., $\alpha = 0.01$, $C = 1$ and $a_0 = 100$ ; (ii) Bottom curve shows the evolution for bounce, i.e., $\alpha = 0.01$, $C = 1$ and $a_0 = -100$.}
\label{Fig1p}
\end{center}
\end{figure}

In Fig. \ref{Fig1p}, we plot the NEC as a function of time, using the exact solution in Eq. (\ref{exactminimal}), for two different initial conditions leading to collapse (top graph) and bounce (bottom graph). While the NEC is perfectly satisfied for a collapse to zero proper volume singularity, usually an indefinite bounce is associated with a violation of NEC. Eventually the bounce leads to a complete dispersal of all the matter distribution inside, as shown in the bottom graph, when $\rho + p \sim 0$. This collapse and dispersal is extremely suggestive of a critical behavior in the system.

\section{Non-minimal Teleparallel Gravity \label{non_min_TG}}
In this section we take a modified teleparallel action where we introduce a nonminimal coupling with a scalar field as follows
\begin{align}
S &= \int d^{4}x e\Bigg[\frac{T}{2\kappa^{2}}
+ \frac{1}{2} \Big(\partial_{\mu}\phi\partial^{\mu}\phi+\xi
T\phi^{2}\Big) - V(\phi)+\mathcal{L}_m\Bigg]\,.\nonumber\\
&\label{action2}
\end{align}

Minimally coupled scalar fields are also popular in the context of TG and such models have been studied quite extensively in the context of cosmic acceleration and reconstruction \cite{Bahamonde:2017ifa,Aslam:2012tj,doi:10.1139/cjp-2012-0281,Bahamonde:2018miw,Geng:2011ka,Xu:2012jf,Jarv:2015odu,Bahamonde:2019shr}. However, it is already established that a minimally coupled scalar field endowed with an interaction potential essentially serves as a fluid distribution, and will therefore do no significant change in outcome as far as the Raychaudhury equation is concerned. A non-minimal coupling on the other hand, inspires an analogy of scalar-geometry interaction in strong gravity limit. Although in the nonminimal case one could use a generalized function of the torsion scalar, we keep the standard $T$ for simplicity.  We also note that the action in Eq.~(\ref{action2}) with the torsion formulation of GR is similar to the standard non-minimal quintessence models of cosmology where the scalar field couples to the Ricci scalar. \\

Variation of action in Eq.~(\ref{action2}) with respect to the tetrad fields yields the equation of motion
\begin{align}
\left(\frac{2}{\kappa^2}+2 \xi
\phi^2 \right)&\left[e^{-1}\partial_{\mu}(ee_A^{\rho}S_{\rho}{}^{\mu\nu} )
-e_{A}^{\lambda}T^{\rho}{}_{\mu\lambda}S_{\rho}{}^{\nu\mu}
-\frac{1}{4}e_{A}^{\nu
}T\right]\nonumber\\
& - e_{A}^{\nu}\left[\frac{1}{2}
\partial_\mu\phi\partial^\mu\phi-V(\phi)\right]+
  e_A^\mu \partial^\nu\phi\partial_\mu\phi \nonumber\\
& + 4\xi e_A^{\rho}S_{\rho}{}^{\mu\nu}\phi
\left(\partial_\mu\phi\right) = e_{A}^{\rho}\Psi_{\rho}{}^{\nu}\,.
\label{eom2}
\end{align}
We now impose the spatially homogeneous geometry of the form (\ref{stellar_collap}) and write the field equations as
\begin{align}
\label{nonminimaleq1}
H^{2} &= \frac{\kappa^2}{3}\Big(\rho_{\phi}+\rho_{m}\Big)\,, \\
\label{nonminimaleq2}
\dot{H} &= -\frac{\kappa^2}{2}\Big(\rho_{\phi}+p_{\phi}+\rho_{m}+p_{m}
\Big)\,,
\end{align}
where the scalar field energy density and pressure is given by
\begin{align}
\label{telerho}
\rho_{\phi} &= \frac{1}{2}\dot{\phi}^{2} + V(\phi)
-  3\xi H^{2}\phi^{2}\,,\\
p_{\phi} &= \frac{1}{2}\dot{\phi}^{2} - V(\phi) +   4 \xi
H \phi\dot{\phi}
 + \xi\left(3H^2+2\dot{H}\right)\phi^2\,. \label{telep}
\end{align}

Using Eqs. (\ref{telerho}) and (\ref{telep}) in Eq. (\ref{rcfinal}), putting $\theta = 3\frac{\dot{a}}{a}$, we write the modified Raychaudhuri equation as
\begin{equation}\label{congruence_nonmin}
\frac{d\theta}{d\tau} = 3 \dot{H} = -\frac{3(\rho_{m} + p_{m} + \dot{\phi}^{2} + 4 \xi H \dot{\phi} \phi)}{2(1 + \xi \phi^2)}.
\end{equation}

Eq. (\ref{congruence_nonmin}) governs the evolution of the time-like congruence, depending on the signature of $\frac{d\theta}{d\tau}$. A negative $\frac{d\theta}{d\tau}$ indicates a collapsing system until $\theta$ reaches $-\infty$, where a zero proper volume singularity forms. However, if $\frac{d\theta}{d\tau}$ changes signature and becomes positive over the course of it's evolution then the collapse of the congruence halts and the geodesics start to move away from each other. Similar to the last section, the onus of avoiding a singularity therefore lies on the behavior of the RHS of Eq. (\ref{congruence_nonmin}).  \\

If we assume that both the energy density $\rho$ and the isotropic pressure $p$ are positive in nature, the nature of the evolution and the predictability depends on the nature of $\xi$ and $\dot{\phi}$. Moreover, $\dot{\phi}^2$ is always positive. If the scalar field increases as a function of time, $\dot{\phi} > 0$ as well. Thus, depending on the signature of $\xi$ the congruence behaves accordingly, for instance if $\xi < 0$ such that $2(1 + \xi \phi^2) < 0$ but $3(\rho_{m} + p_{m} + \dot{\phi}^{2} + 4 \xi H \dot{\phi} \phi) > 0$, somewhere during the collapse, $\dot{H} = \frac{d \theta}{d \tau} > 0$, which means a formation of singularity is avoided. Otherwise there is a formation of zero proper volume singularity. 

For a general coupling function $\xi_{0} \xi(\phi)$ replacing the non-minimal coupling $\xi \phi^2$ in the action, one may generalize the Raychaudhuri equation to write the condition in Eq. (\ref{congruence_nonmin}) as

\begin{equation}\label{congruence_nonmin2}
\frac{d\theta}{d\tau} = -\frac{3(\rho_{m} + p_{m} + \dot{\phi}^{2} + 2\xi_{0}H\dot{\phi} \frac{d\xi}{d\phi})}{2(1 + \xi_{0} \xi(\phi))}\,.
\end{equation}

As an example we present a particular exact solution of the system given by the field equations in Eqs. (\ref{nonminimaleq1}), (\ref{nonminimaleq2}) and the scalar field evolution equation given by
\begin{equation}\label{scalarkg}
\ddot{\phi} + 3H\dot{\phi} + 6\xi H^2 \phi + \frac{dV}{d\phi} = 0\,.
\end{equation}

We solve Eq. (\ref{scalarkg}) by using a theorem on the invertability of these equations \cite{Chakrabarti:2018amp}. The property involves point transforming the equations into an integrable form and is derived from the symmetry analysis of a general classical anharmonic oscillator equation system. The general equation is written as
\begin{equation}\label{anharmonic}
\ddot{\phi}+f_1(t)\dot{\phi}+ f_2(t)\phi+f_3(t)\phi^n=0\,.
\end{equation}
$f_1$, $f_2$ and $f_3$ are unknown functions of some variable, of $t$ at this point, $n$ is a constant. A transformation of this equation into an integrable form requires a pair of point transformations and the condition $n\notin \left\{-3,-1,0,1\right\} $ to be satisfied. Moreover, the coefficients must satisfy the condition
\begin{align}
& \frac{1}{(n+3)}\frac{1}{f_{3}(t)}\frac{d^{2}f_{3}}{dt^{2}} - \frac{(n+4)}{\left( n+3\right) ^{2}}\left[ \frac{1}{f_{3}(t)}\frac{df_{3}}{dt}\right] ^{2}\nonumber\\
& + \frac{(n-1)}{\left( n+3\right) ^{2}}\left[ \frac{1}{f_{3}(t)}\frac{df_{3}}{dt}\right] f_{1}\left( t\right) + \frac{2}{(n+3)}\frac{df_{1}}{dt} \nonumber\\
& + \frac{2\left( n+1\right) }{\left( n+3\right) ^{2}}f_{1}^{2}\left( t\right)=f_{2}(t)\,. \label{criterion1}
\end{align} 
The point transformations are written as 
\begin{align}\label{criterion2}
\Phi\left( T\right) &= C\phi\left( t\right) f_{3}^{\frac{1}{n+3}}\left( t\right)
e^{\frac{2}{n+3}\int^{t}f_{1}\left( x \right) dx } \,,\\
T\left( \phi,t\right) &= C^{\frac{1-n}{2}}\int^{t}f_{3}^{\frac{2}{n+3}}\left(
\xi \right) e^{\left( \frac{1-n}{n+3}\right) \int^{\xi }f_{1}\left( x\right) dx }d\xi \,,\nonumber\\
\end{align}
where $C$ is a constant. Using this property, we solve the scalar field evolution equation assuming it's integrability at the outset. However, this assumption by no means produces unphysical solutions. The scope of this approach has been discussed at length quite recently, in the context of simple scalar field collapse, scalar-Gauss-bonnet gravity and cosmological reconstruction of modified theories of gravity. \\

We assume the potential to be a sum of quadratic and quartic terms of the scalar field, written as
\begin{equation}
V(\phi) = \frac{1}{2} m^2 \phi^2 - \frac{\alpha m^2}{24 f^2}\phi^4\,,
\end{equation}
which is extremely suggestive of a Higgs Potential or an axion dark matter Potential. Using the exact form of the potential, the scalar evolution equation takes the form of
\begin{equation}\label{nonminkg}
\ddot{\phi} + 3H\dot{\phi} + \left(6\xi H^2+m^2\right) \phi - \frac{\alpha m^2}{6 f^2}\phi^3 = 0\,.
\end{equation}
A quick comparison reveals the coefficients to be written as $f_1 = 3 \frac{\dot{a}}{a}$, $f_2 = \left(6\xi H^2+m^2\right)$, $\frac{\alpha m^2}{6 f^2}$ and $n = 3$. Using this, the integrability criterion produces the evolution equation for the radius of the two sphere as
\begin{equation}
\frac{\ddot{a}}{a} + \left(1 - 6\xi\right)\frac{\dot{a}^2}{a^2} - m^2 = 0\,.
\end{equation} 

The first integral of the above equation can be written as
\begin{equation}
\dot{a} = -\left\lbrace a_{0} a^{2(6\xi -1)} + \frac{m^2 a^2}{2 - 6\xi} \right\rbrace^{\frac{1}{2}}\,. 
\end{equation}

\begin{figure}
\begin{center}
\includegraphics[width=0.40\textwidth]{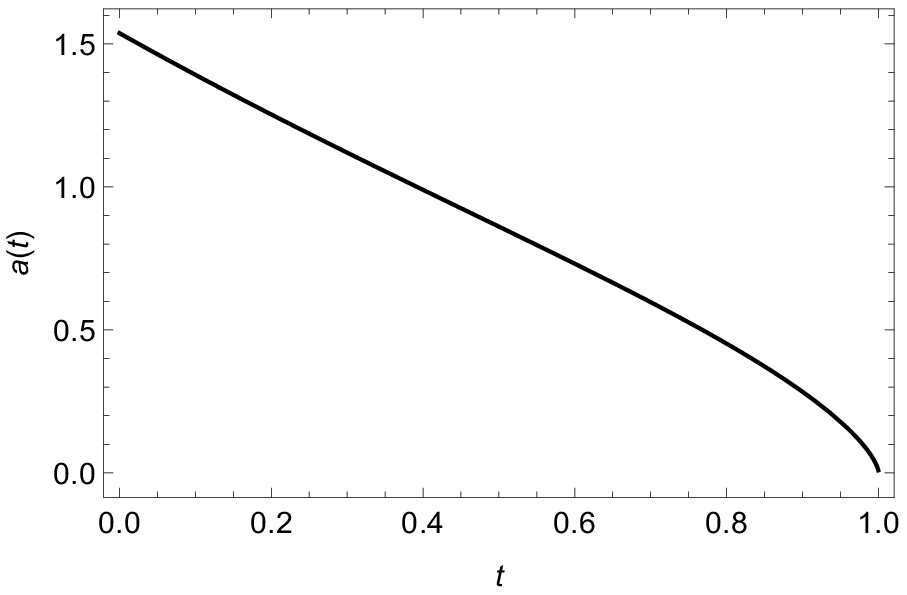}
\includegraphics[width=0.41\textwidth]{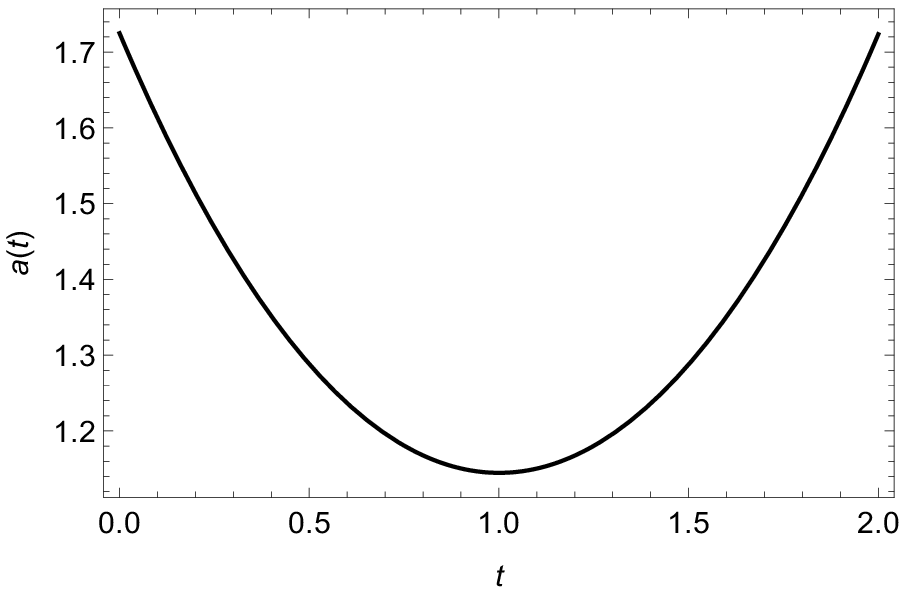}
\caption{Evolution of the radius of two-sphere as given in Eq. (\ref{exactnonminimal}) : (i) Top curve $m = 1$, $C = 1$ and $a_0 = 1$ ; (ii) Bottom curve $m = 1$, $C = 1$ and $a_0 = -1$.}
\label{Fig2}
\end{center}
\end{figure}

As a simple example, we solve the above equation for $\xi = \frac{1}{12}$. This produces the solution for the radius of the two sphere as
\begin{equation}\label{exactnonminimal}
a(t) = \frac{e^{-\frac{\sqrt{2}m(\sqrt{3}t + 3C)}{3}}}{(2m)^{4/3}} \left[e^{3\sqrt{2}mC} - 6 a_{0} m^2 e^{\sqrt{6}m t} \right]^{\frac{2}{3}}\,.
\end{equation}

In Fig. \ref{Fig2}, we plot the evolution as a function of time. The graph on top of the figure shows a plot for $a_0 > 0$. It is clear to note that the collapsing fluid reaches a zero proper volume at a finite future. The time of formation of this zero proper volume singularity may vary depending on the choice of the functional form of the theory, i.e., $\alpha$, however, the qualitatiove behavior remains the same. In the graph below, the evolution is shown for $a_0 < 0$. It is clear that there is no formation of zero proper volume singularity in this case, as the collapsing fluid bounces indefinitely after reaching a minimum cutoff volume. The parameter $a_0$ is a critical parameter of the system whose signature determines the fate of the collapsing system. \\

\begin{figure}
\begin{center}
\includegraphics[width=0.40\textwidth]{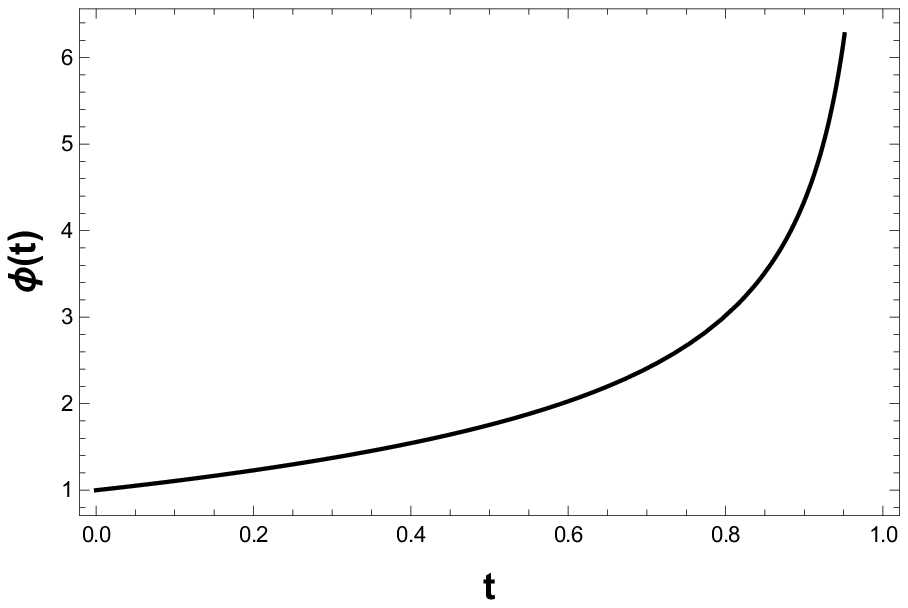}
\includegraphics[width=0.41\textwidth]{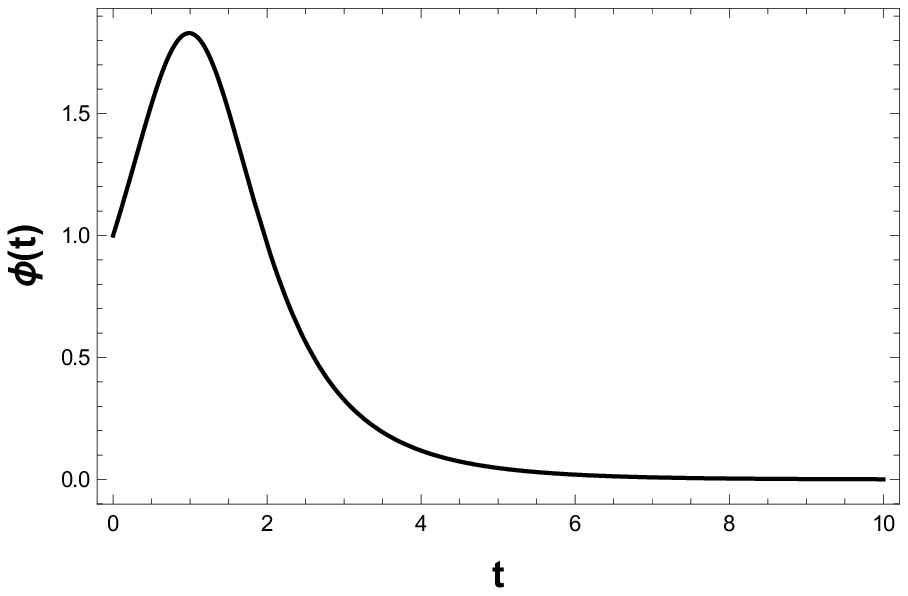}
\caption{Evolution of the scalar field from Eq. (\ref{nonminkg}) : (i) Top curve shows the evolution of the scalar for collapse, i.e., $m = 1$, $C = 1$ and $a_0 = 1$ ; (ii) Bottom curve shows the evolution of the scalar for bounce, i.e., $m = 1$, $C = 1$ and $a_0 = -1$.}
\label{scalarevo}
\end{center}
\end{figure}

We study numerically the evolution of the scalar field of the scalar non-minimal coupling using the Klein Gordon equation as in Eq. (\ref{nonminkg}). The evolution of the scalar field with respect to time is given in Fig. \ref{scalarevo}. It is evident from the figure that, when the sphere collapses onto a zero proper volume, the scalar field diverges around the time of formation of singularity as well. However, from the bottom graph we note that, the collapse and bounce of the sphere is associated with a dispersal of the scalar field to zero value. It may involve radiating or exploding away the strength of scalar field during the indefinite bounce. Using the equation for the NEC, as in Eq. (\ref{nec1}), we also check if the collaping fluid satisfies the NEC. This essentially ensures a positive energy density and that the speed of energy flow of matter is less than the speed of light. 

\begin{figure}
\begin{center}
\includegraphics[width=0.40\textwidth]{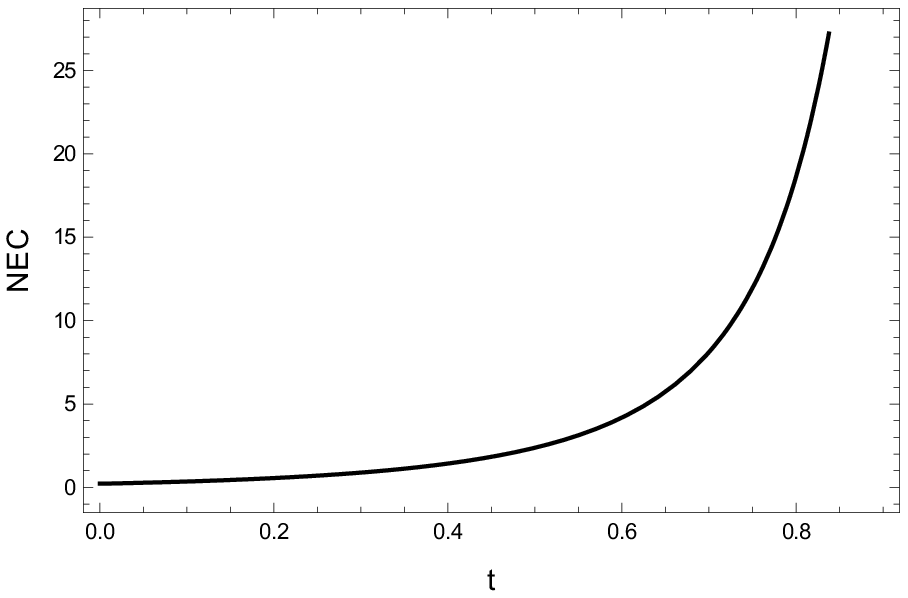}
\includegraphics[width=0.41\textwidth]{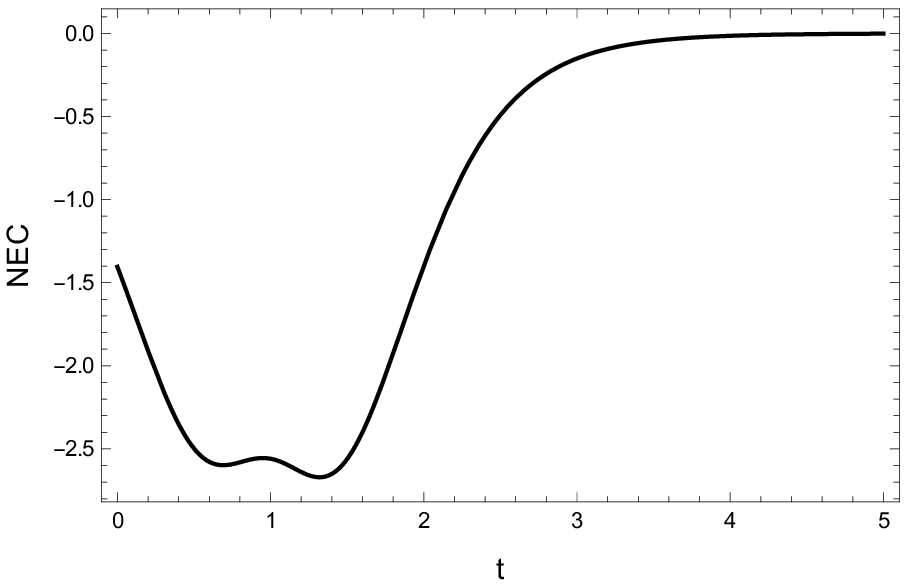}
\caption{Evolution of the NEC as given in Eq. (\ref{nec1}) : (i) Top curve shows the evolution for collapse, i.e., $m = 1$, $C = 1$ and $a_0 = 1$ ; (ii) Bottom curve shows the evolution for bounce, i.e., $m = 1$, $C = 1$ and $a_0 = -1$.}
\label{Fig2p}
\end{center}
\end{figure}

In Fig. \ref{Fig2p}, we plot the NEC as a function of time, using the exact solution in Eq. (\ref{exactnonminimal}), for two different initial conditions leading to collapse (top graph) and bounce (bottom graph). While the NEC is perfectly satisfied for a collapse to zero proper volume singularity, usually an indefinite bounce is associated with a violation of NEC. Eventually the bounce leads to a complete dispersal of all the matter distribution inside, as shown in the bottom graph, when $\rho + p \sim 0$. This collapse and dispersal is extremely suggestive of a critical behavior in the system.

\section{\label{conc} Discussion and Conclusion}
In this work explore TG within the context of stellar collapse through the Raychaudhuri equation and the NEC. TG explores the possibility of replacing the Levi-Civita connection with its Weitzenb\"{o}ck analogue. This has the effect of producing a generically lower-order framework of gravity in which the metric is exchanged with the tetrad in terms of the fundamental dynamical object of the theory. We use the relations between TG and standard gravity to relate the components of the Ricci tensor with their theory-dependent teleparallel analogues through Eq.(\ref{Ricci_ten_comp}). The effect of this is that the Raychaudhuri equation in Eq.(\ref{Raych-time}) can be used to determine the congruence of neighbouring particle geodesics. The Raychaudhuri equation is a general result for Riemann manifolds which is why we can use it in this context. We then use the NEC to determine which of these solutions indeed produces collapsing models.

The scenario of a spatially homogeneous collapsing stellar interior is investigated in \S.~\ref{collapse} within the $f(T)$ gravity extension to TEGR. Here, we assume the same rational as the widely popular $f(\mathring{R})$ gravity framework. By probing this scenario of TG with a perfect fluid, we find the condition for stellar collapse, both in terms of the strightforward torsion scalar but also as a function of $H(t)$. Then by using the NEC, we determine trial solutions that satisfy this condition. As we show in this section, these models are consistent with a number of literature proposals, and moreover are instrumental in determining general conditions for collapse within $f(T)$ gravity.

We again consider this scenario in \S.\ref{non_min_TG}, but in this case we take consider an interesting nonminimally coupled scalar field which is added to the TEGR Lagrangian in the action in Eq.(\ref{action2}). In this case, we need to also use a very intrigueing theorem within calculus on solutions of anharmonic oscillator systems in Eq.(\ref{anharmonic}). In this setup, we can then find solutions to the evolution equations. In this part of the work, we plot the NEC to determine when this is satisfied in Fig.(\ref{Fig2p}). Scalar-tensor theories are very interesting in TG due to its organically lower-order nature which produces a much wider range of models that remain second-order than their standard gravity analogues.

TG has been mainly studies in cosmology and so works in stellar systems can reveal a lot about the physically viable theories from cosmology. Collapse models provide an intriguing test bed in which to perform these studies and may elucidate several literature models within their strong field regime.

\bibliographystyle{unsrt}

\begin{thebibliography}{10}

\bibitem{misner1973gravitation}
C. W. Misner, K. S. Thorne, and J. A. Wheeler.
\newblock Number pt. 3 in Gravitation. W. H. Freeman, 1973.

\bibitem{Clifton:2011jh}
T. Clifton, P. Ferreira, A. Padilla, and C. Skordis.
\newblock {\em Phys. Rept.}, 513 : 1, 2012.

\bibitem{Baudis:2016qwx}
L. Baudis.
\newblock {\em J. Phys.}, G43(4):044001, 2016.

\bibitem{Bertone:2004pz}
G. Bertone, D. Hooper, and J. Silk.
\newblock {\em Phys. Rept.}, 405:279--390, 2005.

\bibitem{Peebles:2002gy}
P. J. E. Peebles and B. Ratra.
\newblock {\em Rev. Mod. Phys.}, 75 : 559, 2003.
\newblock [,592(2002)].

\bibitem{Copeland:2006wr}
E. J. Copeland, M. Sami, and S. Tsujikawa.
\newblock {\em Int. J. Mod. Phys.}, D15 : 1753, 2006.

\bibitem{Riess:1998cb}
A. G. Riess et. al.
\newblock {\em Astron. J.}, 116 : 1009, 1998.

\bibitem{Perlmutter:1998np}
S. Perlmutter et. al.
\newblock {\em Astrophys. J.}, 517 : 565, 1999.

\bibitem{RevModPhys.61.1}
S. Weinberg.
\newblock {\em Rev. Mod. Phys.}, 61 : 1, 1989.

\bibitem{Gaitskell:2004gd}
R. J. Gaitskell.
\newblock {\em Ann.\ Rev.\ Nucl.\ Part.\ Sci.}, 54 : 315, 2004.

\bibitem{Riess:2019cxk}
A. G. Riess, S. Casertano, W. Yuan, L. M. Macri, and D. Scolnic.
\newblock {\em Astrophys. J.}, 876(1) : 85, 2019.

\bibitem{Wong:2019kwg}
K. C. Wong et. al.
\newblock {arXiv:1907.04869v2 [astro-ph.CO]}.

\bibitem{Aghanim:2018eyx}
N. Aghanim et. al.
\newblock { arXiv:1807.06209v2 [astro-ph.CO] }.

\bibitem{Ade:2015xua}
P. A. R. Ade et. al.
\newblock {\em Astron. Astrophys.}, 594 : A13, 2016.

\bibitem{Freedman:2019jwv}
W. L. Freedman et. al.
\newblock {\em The Astrophysical Journal}, 882 : 34, 2019.

\bibitem{Baker:2019nia}
J. Baker et. al.
\newblock {arXiv:1907.06482v2 [astro-ph.IM]}.

\bibitem{2017arXiv170200786A}
P. {Amaro-Seoane} et~al.
\newblock {\em arXiv e-prints}, page arXiv : 1702.00786.

\bibitem{Graef:2018fzu}
L. L. Graef, M. Benetti, and J. S. Alcaniz.
\newblock {\em Phys. Rev. D}, 99(4) : 043519, 2019.

\bibitem{Abbott:2017xzu}
B.P. Abbott et. al.
\newblock {\em Nature}, 551(7678) : 85, 2017.

\bibitem{Capozziello:2011et}
S. Capozziello and M. De~Laurentis.
\newblock {\em Phys. Rept.}, 509 : 167, 2011.

\bibitem{Sotiriou:2008rp}
T. P. Sotiriou and V. Faraoni.
\newblock {\em Rev. Mod. Phys.}, 82 : 451, 2010.

\bibitem{Faraoni:2008mf}
V. Faraoni.
\newblock {\em  arXiv:0810.2602v1 [gr-qc]}.

\bibitem{BeltranJimenez:2019tjy}
J. Beltrán Jiménez, L. Heisenberg, and T. S. Koivisto.
\newblock {\em Universe}, 5(7):173, 2019.

\bibitem{nakahara2003geometry}
M.~Nakahara.
\newblock Graduate student series in physics. Taylor \& Francis, 2003.

\bibitem{Aldrovandi:2013wha}
R. Aldrovandi and J. G. Pereira.
\newblock Springer, Dordrecht, 2013.

\bibitem{Cai:2015emx}
Y. F. Cai, S. Capozziello, M. De~Laurentis, and E. N. Saridakis.
\newblock {\em Rept. Prog. Phys.}, 79(10):106901, 2016.

\bibitem{Krssak:2018ywd}
M.~Kr\v{s}\v{s}\'{a}k, R.~J. van~den Hoogen, J.~G. Pereira, C.~G. Böhmer, and
  A.~A. Coley.
\newblock {\em Class. Quant. Grav.}, 36(18):183001, 2019.

\bibitem{Weitzenbock1923}
R.~Weitzenb\"{o}ock.
\newblock {\em `Invariantentheorie'}.
\newblock Noordhoff, Gronningen, 1923.

\bibitem{Lovelock:1971yv}
D.~Lovelock.
\newblock {\em J. Math. Phys.}, 12 : 498, 1971.

\bibitem{Gonzalez:2015sha}
P.~A. Gonzalez and Y. Vasquez.
\newblock {\em Phys. Rev.}, D92(12):124023, 2015.

\bibitem{Bahamonde:2019shr}
S. Bahamonde, K. F. Dialektopoulos, and J. L. Said.
\newblock {\em Phys. Rev.}, D100(6):064018, 2019.

\bibitem{Blixt:2018znp}
D. Blixt, M. Hohmann, and C. Pfeifer.
\newblock {\em Phys.Rev.D}, 99(8):084025, 2019.

\bibitem{Blixt:2019mkt}
D. Blixt, M. Hohmann, and C. Pfeifer.
\newblock {\em Universe}, 5(6):143, 2019.

\bibitem{Aldrovandi:2004fy}
R. Aldrovandi, J. G. Pereira, and K. H. Vu.
\newblock In {\em The Tenth Marcel Grossmann Meeting}, page 1505, 2006.

\bibitem{Ferraro:2006jd}
R. Ferraro and F. Fiorini.
\newblock {\em Phys. Rev.}, D75:084031, 2007.

\bibitem{Ferraro:2008ey}
R. Ferraro and F. Fiorini.
\newblock {\em Phys. Rev.}, D78:124019, 2008.

\bibitem{Bengochea:2008gz}
G. R. Bengochea and R. Ferraro.
\newblock {\em Phys. Rev.}, D79:124019, 2009.

\bibitem{Linder:2010py}
E. V. Linder.
\newblock {\em Phys. Rev.}, D81:127301, 2010.
\newblock [Erratum: Phys. Rev.D82,109902(2010)].

\bibitem{Chen:2010va}
S. H. Chen, J. B. Dent, S. Dutta, and E. N. Saridakis.
\newblock {\em Phys. Rev.}, D83:023508, 2011.

\bibitem{Bahamonde:2019zea}
S. Bahamonde, K. Flathmann, and C. Pfeifer.
\newblock {\em Phys.\ Rev.\ D}, 100(8):084064, 2019.

\bibitem{Nesseris:2013jea}
S.~Nesseris, S.~Basilakos, E.~N. Saridakis, and L.~Perivolaropoulos.
\newblock {\em Phys. Rev.}, D88:103010, 2013.

\bibitem{Farrugia:2016qqe}
G. Farrugia and J. L. Said.
\newblock {\em Phys. Rev.}, D94(12):124054, 2016.

\bibitem{Finch:2018gkh}
A. Finch and J. L. Said.
\newblock {\em Eur.Phys.J.C}, 78(7):560, 2018.

\bibitem{Farrugia:2016xcw}
G. Farrugia, J. L. Said, and M. L. Ruggiero.
\newblock {\em Phys. Rev.}, D93(10):104034, 2016.

\bibitem{Iorio:2012cm}
L. Iorio and E. N. Saridakis.
\newblock {\em Mon. Not. Roy. Astron. Soc.}, 427:1555, 2012.

\bibitem{Ruggiero:2015oka}
M. L. Ruggiero and N. Radicella.
\newblock {\em Phys. Rev.}, D91:104014, 2015.

\bibitem{Deng:2018ncg}
X. M. Deng.
\newblock {\em Class.Quant.Grav.}, 35(17):175013, 2018.

\bibitem{Bahamonde:2015zma}
S. Bahamonde, C. G. Böhmer, and M. Wright.
\newblock {\em Phys. Rev.}, D92(10):104042, 2015.

\bibitem{Capozziello:2018qcp}
S. Capozziello, M. Capriolo, and M. Transirico.
\newblock {\em  International Journal of Geometric Methods in Modern PhysicsVol. 15, No. supp01, 1850164 (2018)}.

\bibitem{Bahamonde:2016grb}
S. Bahamonde and S. Capozziello.
\newblock {\em Eur. Phys. J.}, C77(2):107, 2017.

\bibitem{Paliathanasis:2017flf}
A. Paliathanasis.
\newblock {\em JCAP}, 1708(08):027, 2017.

\bibitem{Farrugia:2018gyz}
G. Farrugia, J. L. Said, V. Gakis, and E. N. Saridakis.
\newblock {\em Phys. Rev.}, D97(12):124064, 2018.

\bibitem{Bahamonde:2016cul}
S. Bahamonde, M. Zubair, and G. Abbas.
\newblock {Thermodynamics and cosmological reconstruction in $f(T,B)$ gravity}.
\newblock {\em Phys. Dark Univ.}, 19 : 78, 2018.

\bibitem{Wright:2016ayu}
M. Wright.
\newblock {\em Phys. Rev.}, D93(10):103002, 2016.

\bibitem{Bahamonde:2017ifa}
S. Bahamonde.
\newblock {\em Eur. Phys. J.}, C78(4):326, 2018.

\bibitem{Harko:2014aja}
T. Harko, F. S. N. Lobo, G. Otalora, and E. N. Saridakis.
\newblock {\em JCAP}, 1412:021, 2014.

\bibitem{Pace:2017aon}
M. Pace and J. L. Said.
\newblock {\em Eur. Phys. J.}, C77(2):62, 2017.

\bibitem{Farrugia:2016pjh}
G. Farrugia and J. L. Said.
\newblock {\em Phys. Rev.}, D94(12):124004, 2016.

\bibitem{Saez-Gomez:2016wxb}
D. Saez-Gomez, C. Sofia Carvalho, F. S. N. Lobo, and I. Tereno.
\newblock {\em Phys. Rev.}, D94(2):024034, 2016.

\bibitem{Momeni:2014jja}
D. Momeni and R. Myrzakulov.
\newblock {\em Int. J. Geom. Meth. Mod. Phys.}, 11(08):1450077, 2014.

\bibitem{Nassur:2015zba}
S.~B. Nassur, M.~J.~S. Houndjo, A.~V. Kpadonou, M.~E. Rodrigues, and J.~Tossa.
\newblock {\em Astrophys. Space Sci.}, 360(2):60, 2015.

\bibitem{Pace:2017dpu}
M. Pace and J. Levi Said.
\newblock {\em Eur. Phys. J.}, C77(5):283, 2017.

\bibitem{1955PhRv...98.1123R}
A. {Raychaudhuri}.
\newblock {\em Physical Review}, 98(4):1123, 1955.

\bibitem{hawking_ellis_1973}
S.~W. Hawking and G.~F.~R. Ellis.
\newblock Cambridge Monographs on Mathematical Physics. Cambridge University
  Press, 1973.

\bibitem{Harko:2012ve}
T. Harko and F. S. N. Lobo.
\newblock {\em Phys. Rev. D}, 86:124034, 2012.

\bibitem{aldrovandi1995introduction}
R.~Aldrovandi and J. G. Pereira.
\newblock World Scientific, 1995.

\bibitem{Li:2010cg}
B. Li, T. P. Sotiriou, and J. D. Barrow.
\newblock {\em Phys. Rev.}, D83:064035, 2011.

\bibitem{Krssak:2015oua}
M. Kr\v{s}\v{s}\'{a}k and E. N. Saridakis.
\newblock {\em Class. Quant. Grav.}, 33(11):115009, 2016.

\bibitem{Tamanini:2012hg}
N. Tamanini and C. G. Boehmer.
\newblock {\em Phys. Rev.}, D86:044009, 2012.

\bibitem{Bahamonde:2017wwk}
S. Bahamonde, C. G. Böhmer, and M. Kr\v{s}\v{s}\'{a}k.
\newblock {\em Phys. Lett.}, B775:37, 2017.

\bibitem{RevModPhys.48.393}
F. W. Hehl, P. von~der Heyde, G.~David Kerlick, and J. M. Nester.
\newblock {\em Rev. Mod. Phys.}, 48:393, Jul 1976.

\bibitem{Aldrovandi:2004db}
R.~Aldrovandi, P.~B. Barros, and J.~G. Pereira.
\newblock { arXiv:gr-qc/0402022v2}.

\bibitem{Koivisto:2019jra}
T. Koivisto, M. Hohmann, and L. Marzola.
\newblock {arXiv:1909.10415v1 [gr-qc]}.

\bibitem{Abedi:2017jqx}
H. Abedi and S. Capozziello.
\newblock {\em Eur. Phys. J.}, C78(6):474, 2018.

\bibitem{Farrugia:2020fcu}
G. Farrugia, J. L. Said, and A. Finch.
\newblock {\em Universe}, 6:34, 2020.

\bibitem{Kolassis_1988}
C. A Kolassis, N. O. Santos, and D. Tsoubelis.
\newblock {\em Classical and Quantum Gravity}, 5(10):1329--1338, oct 1988.

\bibitem{2014PhRvD..90h4011G}
R. {Goswami}, A. M. {Nzioki}, S.~D. {Maharaj}, and S. G. {Ghosh}.
\newblock {\em \prd}, 90(8):084011, 2014.

\bibitem{Aslam:2012tj}
A. Aslam, M. Jamil, D. Momeni, and R. Myrzakulov.
\newblock {\em Can. J. Phys.}, 91:93--97, 2013.

\bibitem{doi:10.1139/cjp-2012-0281}
A. Aslam, M. Jamil, D. Momeni, and R. Myrzakulov.
\newblock {\em Can. J. Phys.}, 91:93--97, 2013.

\bibitem{Bahamonde:2018miw}
S. Bahamonde, M. Marciu, and P. Rudra.
\newblock {\em JCAP} 04 (2018) 056.

\bibitem{Geng:2011ka}
C. Q. Geng, C. C. Lee, and E. N. Saridakis.
\newblock {\em JCAP}, 1201:002, 2012.

\bibitem{Xu:2012jf}
C. Xu, E. N. Saridakis, and G. Leon.
\newblock {\em JCAP}, 1207:005, 2012.

\bibitem{Jarv:2015odu}
L. Jarv and A. Toporensky.
\newblock {\em Phys. Rev.}, D93(2):024051, 2016.

\bibitem{Chakrabarti:2018amp}
S. Chakrabarti, K. Bamba, and J. L. Said.
\newblock {\em Gen. Rel. Grav.}, 52(1):7, 2020.

\end{thebibliography}

\end{document}